\documentclass{article}
\usepackage{spconf,amsmath,graphicx, multirow, makecell}
\ninept

\title{TWO-PATH GMM-RESNET AND GMM-SENET FOR ASV SPOOFING DETECTION}
\name{Zhenchun Lei, Hui Yan, Changhong Liu, Minglei Ma, Yingen Yang}
\address{School of Computer and Information Engineering, Jiangxi Normal University, Nanchang, China}
\begin{document}
%
\maketitle
\begin{abstract}
The automatic speaker verification system is sometimes vulnerable to various spoofing attacks. The 2-class Gaussian Mixture Model classifier for genuine and spoofed speech is usually used as the baseline for spoofing detection. However, the GMM classifier does not separately consider the scores of feature frames on each Gaussian component. In addition, the GMM accumulates the scores on all frames independently, and does not consider their correlations. We propose the two-path GMM-ResNet and GMM-SENet models for spoofing detection, whose input is the Gaussian probability features based on two GMMs trained on genuine and spoofed speech respectively. The models consider not only the score distribution on GMM components, but also the relationship between adjacent frames. A two-step training scheme is applied to improve the system robustness. Experiments on the ASVspoof 2019 show that the LFCC+GMM-ResNet system can relatively reduce min-tDCF and EER by 76.1\% and 76.3\% on logical access scenario compared with the GMM, and the LFCC+GMM-SENet system by 94.4\% and 95.4\% on physical access scenario. After score fusion, the systems give the second-best results on both scenarios.
\end{abstract}
\section{Introduction}
Automatic Speaker Verification (ASV) \cite{8461375} systems have shown good performances thanks to the high capacity of the deep learning models. However, the ASV technologies are still vulnerable to some real-life spoofing attacks \cite{WU2015130}, such as mimicry \cite{g20_interspeech}, text-to-speech \cite{sahidullah15_interspeech}, voice conversion \cite{9262021}, and replay \cite{9413828}.

Various spoofing technologies have been proposed to improve detection performance, and they can be roughly grouped into two categories, i.e., the feature engineering methods and the classification methods. The feature engineering methods focuse on designing discriminative features \cite{sahidullah15_interspeech}, such as LFCC \cite{6712706} and CQCC \cite{TODISCO2017516}. On the other hand, the classification methods employ more and more neural network models for speech spoofing detection. The most commonly used models are Convolutional Neural Networks (CNN), which have shown remarkable performance in the task of spoofing detection. Lavrentyeva et al. \cite{lavrentyeva17_interspeech} employed Light Convolutional Neural Networks (LCNN) with max filter map activation function and achieved the best performance in ASVspoof 2017 challenge. Alanis et al. \cite{gomezalanis19_interspeech} proposed a hybrid LCNN plus RNN architecture which combines the ability of the LCNNs for extracting discriminative features at frame level with the capacity of Gated Recurrent Unit (GRU) based RNNs for learning long-term dependencies of the subsequent deep features. Lavrentyeva et al. \cite{lavrentyeva19_interspeech} investigated the efficiency of angular margin based softmax activation for training robust deep LCNN classifier. Wu et al. \cite{wu20c_interspeech} proposed a novel method referred to as feature genuinization that learns a transformer with CNN using the characteristics of only genuine speech, and then use this genuinization transformer with a LCNN classifier for detection of synthetic speech attacks.

A practical challenge in deep convolutional networks is the vanishing gradients that makes it hard for lower-layers to receive useful update signals during training. He et al. \cite{7780459} proposed an effective solution called residual networks (ResNet) which employ skip connections that act as shortcuts allowing training updates to back-propagate faster towards the lower layers. The ResNet model can achieve better classification performance and has achieved great success in spoofing detection. Alzantot et al. \cite{alzantot19_interspeech} proposed deep residual neural networks for audio spoofing detection, which process MFCC, CQCC and spectrogram input features, respectively. Lai et al. \cite{8682640} proposed attentive filtering network, which is composed of an attention-based filtering mechanism that enhances feature representation in both frequency and time domains, and a ResNet-based classifier. ASSERT system \cite{lai19b_interspeech} is a pipeline for DNN-base approach, which is based on variants of Squeeze-Excitation Network \cite{8701503} and ResNet with statistical pooling to address anti-spoofing. Parasu et al. \cite{parasu20_interspeech} proposed a novel Light-ResNet architecture that provides good generalization across databases and attack types. Tak et al. \cite{9414234} applied RawNet2 to anti-spoofing, which ingests raw audio and has potential to learn cues that are not detectable using more traditional countermeasure solutions.

The classical Gaussian Mixture Model (GMM) is generally used as the baseline system for spoofing detection. In the GMM, the final scores are accumulated on all feature frames independently, and the Gaussian component score distribution information is discarded. The relationship between adjacent frames along the time has been ignored, and the frame feature correlations are not considered. In our previous papers, we use the conventional CNN model \cite{lei20_interspeech} and a simple Transformer model \cite{10.1007/978-3-030-86608-2_50} with the Gaussian probability feature, respectively. In this paper, we employ the ResNet and SENet which can get the better performance. On the other hand, we propose the two-step training scheme and the two-path GMM-ResNet and GMM-SENet for speech spoofing detection in this paper.

\section{log Gaussian probability feature}
The Gaussian Mixture Model (GMM) is a weighted sum of $M$ Gaussian probability density functions with different mixing weights and different parameters, namely, means and covariance matrices. For a feature vector, $x\in R^{D\times 1}$, the mixture density used for the likelihood function has the following form:
\begin{equation}
	p(x)=\sum_{i=1}^M w_ip_i(x) \label{eq_gmm}
\end{equation}
The density is a weighted linear combination of $M$ unimodal Gaussian densities. $w_i$ is the $ith$ Gaussian component weight, and $p_i(x)$ is parameterized by a mean $D\times 1$ vector, $\mu_i$, and a $D\times D$ covariance matrix, $\Sigma_i$:
\begin{equation}
	p_i (x)=\frac{1}{(2\pi)^{D/2} |\Sigma_i|^{1/2} } exp\{ -\frac{1}{2}(x-\mu_i)'\Sigma_i^{-1} (x-\mu_i) \} \label{eq_gm}
\end{equation}
In order for $p(x)$ to be a probability density function with integral 1 over the input space, the additional constraints on the weights of the mixture must hold:
\begin{equation}
	\sum_{i=1}^{M} w_i =1, 
	w_i \ge 0
\end{equation}

The estimation of the parameters of the model is carried out by the Expectation Maximization (EM) algorithm, aiming at maximizing the likelihood of a set of samples drawn independently from the unknown density.
In ASV spoofing detection field, the GMM based system is general used as the baseline system, which contains two GMMs: one for genuine speech and one for spoofed speech. For a given test speech utterance $X=\left\lbrace x_1,x_2,...,x_N\right\rbrace $, the log-likelihood ratio is used to make the genuine/spoofed decision, and the log-likelihood ratio is defined in the following formula:
\begin{equation}
	score_{gmm}=\log p(X|\lambda_g) - \log p(X|\lambda_s)
\end{equation}
where $\lambda_g$ and $\lambda_s$ are the GMMs for genuine and spoofed speech respectively.
For a feature vector, the GMM sums the $M$ Gaussian probability density values. But the GMM does not consider the contribution of every Gaussian component to the final score separately. The distributions of genuine and spoofed speech are different in feature space, and their score distributions on all Gaussian components are also different. So, this score distribution information is useful for spoofing detection.
For a raw frame feature $x$ (CQCC or LFCC in our experiments), the size of new feature $y$ is the order of GMM and the element $y_i$ is:
\begin{align}
	y_i&=\log p_i(x) \nonumber \\
	~ &= -\frac{D}{2} \log(2\pi)-\frac{1}{2} \log(|\Sigma_i|) - \frac{1}{2}(x-\mu_i)'\Sigma_i^{-1} (x-\mu_i) \nonumber \\
	~ &= -\frac{1}{2}x'\Sigma_i^{-1}x + x'\Sigma_i^{-1}\mu_i + C
\end{align}
where $C$ is constant. After that, the mean $mean_{y_i}$ and standard deviation $std_{y_i}$ of $y_i$ on the whole training data are calculated, and the mean and variance normalization is applied:
\begin{equation}
	f_i=\frac{y_i-mean_{y_i}}{std_{y_i}}
\end{equation}
In order to simply the calculation, the constant $C$ can be removed. So, the Log Gaussian Probability (LGP) feature can be written as:
\begin{equation}
	y_i'=-\frac{1}{2}x'\Sigma_i^{-1}x + x'\Sigma_i^{-1}\mu_i
\end{equation}
After normalization, the LGP feature is:
\begin{equation}
	f_i'=\frac{y_i'-mean_{y_i'}}{std_{y_i'}}
\end{equation}

The LGP feature is the log Gaussian probability of the raw feature on each Gaussian component. Its physical meaning is the weighted distance between the frame feature and the center of each Gaussian component in the feature space.

\section{GMM-ResNet and GMM-SENet}
For a speech feature sequence, the GMM accumulates the scores on all frames independently and ignores the relationship between adjacent frames. The context of frame is helpful to improve the performance for spoofing detection.

\subsection{GMM-ResNet and GMM-SENet}
Convolutional Neural Networks (CNN) based models have been popularly applied in various machine learning tasks. Many neural network architectures have the problem of “vanishing gradients” with the increase of the number of layers. The ResNet \cite{7780459} architecture is designed to solve this problem with the idea of shortcut connections. They make it possible to guarantee that with the increase in the number of layers of a neural network it will not need to learn the identical transformation in order to remain no worse than its counterpart with fewer layers. This is due to the fact that we immediately add a direct connection between the output of each layer with the input of the next to it layer. ResNet has been successfully applied in many fields, and we proposed GMM-ResNet and GMM-SENet models for speech spoofing detection. The proposed models consider not only the frame scores on GMM components, but also the local relationship between adjacent frames. Table \ref{tab:gmm_resnet} shows their architectures.
\begin{table}[t]\setlength\tabcolsep{3pt} 
	\begin{center}
		\caption{GMM-ResNet and GMM-SENet architectures. Numbers denoted in Conv1d refer to kernel size, stride, and number of filters.} \label{tab:gmm_resnet}
		\begin{tabular}{ccc}
			\hline
			Layer & Input: CQCC or LFCC & Output shape  \\
			\hline
			GMM & 512 & (512, 400 or 1000) \\
			\hline
			Conv layer & \makecell[c]{Conv1d(3,1,512) \\ BN \\ ReLU} & (512, 400 or 1000) \\
			\hline
			Res blocks & 
			$
			\begin{pmatrix}
				Conv1d(3,1,512) \\ BN \\ ReLU \\ Conv1d(3,1,512) \\ BN \\ ReLU \\   ---  \\ SE
			\end{pmatrix} \times 6
			$
			& (512, 400 or 1000)  \\
			\hline
			Max pool & AdaptiveMaxPool1d  & (512)  \\
			\hline
			FC & 512 & (2) \\
			\hline
		\end{tabular}
	\end{center}
\end{table}

The GMM-ResNet model use CQCC or LFCC as input feature, which is truncated or repeated such that all utterances are of the same length (400 in LA and 1000 in PA) and inputted to the GMM trained on the whole training data set. After the Gaussian probability feature is extracted, one convolutional layer and 6 residual blocks are applied. Then a max-overtime pooling operation is applied over the feature map and the maximum value is taken as the embedding vector. The idea is to capture the most important feature - one with the highest value. Finally, the embedding vector is inputted to the fully connected layer for spoofing detection.

Squeeze-Excitation Network (SENet) \cite{8701503} is a building block which recalibrates channel-wise responses adaptively by explicitly modeling the relationship between channels. It can be directly added to existing residual block, and has been shown significant performance improvements. The GMM-SENet is the extension of GMM-ResNet in which SE block is integrated into each residual block.

The Unified Feature Map (UFM) \cite{9053303} is used for the different lengths of evaluation utterances. We first extended all utterances to multiple of $N$ frames, and the extended feature map was broken down into segments with length $N$ frames and overlap $N/2$ frames. The score of utterance is computed by averaging the scores over all segments.

\subsection{Two-path GMM-ResNet and GMM-SENet}
The Gaussian probability feature depend on the GMM. There are two GMMs in the baseline system, and we can get two types of Gaussian probability features. So, we propose a two-path architecture with ResNet or SENet.
\begin{figure}[htb]
		\centerline{\includegraphics[width=8.5cm]{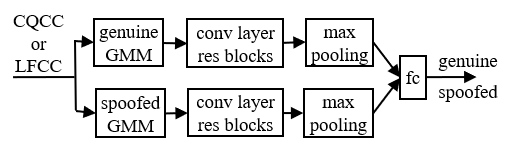}}
	\caption{Diagram of the two-path GMM-ResNet or GMM-SENet.}
	\label{fig:res}
\end{figure}

The two-path GMM-ResNet contains two identical ResNets, each of which has the same architecture. The input of convolutional network is the Gaussian probability feature extracted separately on two GMMs, which are trained on genuine and spoofed speech respectively in the baseline system. The two-path embedding vectors are concatenated and inputted to the fully connected layer for spoofing detection. The two-path GMM-SENet architecture is similar except for SE residual blocks.

\subsection{Two-step training scheme}
The multi-step training scheme \cite{jung18b_interspeech} is usually used to avoid overfitting problem. In \cite{jung20c_interspeech}, the training scheme first trains a CNN and then expands to a CNN-GRU, and demonstrates significant improvement. Considering the large number of network parameters, we also use a two-step training scheme aiming at improving the robustness. In the first step, the conv layer and res blocks are trained. A fully connected layer with softmax output is temporarily attached to the max-pool layer in each path for spoofing detection. The conv layer and res blocks are trained in the same as that of one-path model, and the cross-entropy loss function is used. In the second step, the temporary layers are removed, and the whole architecture is applied. The conv layer and res blocks are firstly pretrained with additional classifiers independently, and then we freeze the model parameters of conv and res blocks to train the system classifier. 

\section{Experiments}
\subsection{Experiment Settings}
The proposed models are evaluated on the ASVspoof 2019 \cite{todisco19_interspeech} database, which contains two partitions for the assessment of logical access (LA) and physical access (PA) scenarios. The LA scenario involves spoofing attacks that are injected directly into the ASV system, and the PA scenario is assumed to capture speech data by a microphone in a physical, reverberant space. ASVspoof 2019 corpus is divided into three no speakers overlap subsets: training, development and evaluation. Performance is measured in terms of minimum tDCF and EER.

The LFCC and CQCC are used as acoustic features in our experiments. The feature extractors are implementation of spoofing detection baseline system provided by the organizers with the default configuration. The GMM has 512 mixtures, and has been trained by 30 EM iterations using the MSR Identity Toolbox \cite{sadjadi2013msr} implementation.

We have implemented our neural network models with PyTorch and tested them on our workstation with a 64-bit Linux system, Xeon processor and a GTX 2080 TI graphics card. Cross-entropy loss is adopted as the loss criterion and Adam optimizer with learning rate of 0.0001 is used during the training process. The batch size is set to 32 and the number of epochs is 100 in all experiments.

\subsection{Results on ASVspoof 2019 LA scenario}
The LA scenario contains bona fide speech and spoofed speech data generated by 17 different TTS and VC systems. Table \ref{tab:asvspoof2019la} shows the results on the ASVspoof 2019 LA scenario obtained by the baseline systems and our models with LFCC and CQCC feature.
\begin{table}[t]\setlength\tabcolsep{3pt} 
	\begin{center}
		\caption{Results on ASVspoof 2019 LA scenario in terms of minimum tDCF and EER (\%). 2P refers to the two-path architecture and 2S refers to the two-step training scheme.} \label{tab:asvspoof2019la}
		\begin{tabular}{clcccc}
			\hline
			\multirow{2}*{Feature} & \multirow{2}*{Model}  & \multicolumn{2}{c}{Dev}  & \multicolumn{2}{c}{Eval} \\
			\cline{3-6}
		    &	& tDCF & EER  & tDCF & EER \\
			\hline
			\multirow{8}*{CQCC}&GMM \cite{todisco19_interspeech} & 0.0123 & 0.43 & 0.2366 & 9.57 \\
			&GMM(ours)          &0.0066  &0.24  &0.2135 &8.97 \\
			\cline{2-6}
			&ResNet(without GMM)   &0.0105	&0.31	&0.3840	&12.37\\
			&GMM-ResNet	        &0.0016	 &0.12	&0.2054	&8.24 \\
			&GMM-ResNet(2P)  	&0.0014	 &0.04	&0.2400	&8.52 \\
			&GMM-ResNet(2P2S)	&0.0024  &0.09	&0.2039	&7.83 \\
			\cline{2-6}
			&SENet(without GMM)       	&0.0071	&0.24	&0.3636	&11.83 \\
			&GMM-SENet       	&0.0003  &0.03	&0.2567	&8.70 \\
			&GMM-SENet(2P)   	&0.0023  &0.08	&0.2248	&8.79 \\
			&GMM-SENet(2P2S)	&0.0001  &0.01	&0.2000	&7.67 \\
			\hline
			\multirow{8}*{LFCC}&GMM\cite{todisco19_interspeech} & 0.0663 & 2.71 & 0.2116 & 8.09 \\
			&GMM(ours)          &0.1031 &3.77   &0.2084 &7.59 \\
			\cline{2-6}
			&ResNet(without GMM)   &0.0346	&1.34	&0.1335	&5.29\\
			&GMM-ResNet	        &0.0119	&0.43	&0.0538	&2.06 \\
			&GMM-ResNet(2P)   	&0.0104	&0.35	&0.0557	&2.02 \\
			&GMM-ResNet(2P2S)	&0.0094	&0.35	&\textbf{0.0498}	&\textbf{1.80} \\
			\cline{2-6}
			&SENet(without GMM)       	&0.0273	&1.06	&0.0946	&3.87 \\
			&GMM-SENet	        &0.0129	&0.43	&0.0866	&3.20 \\
			&GMM-SENet(2P)   	&0.0113	&0.38	&0.0922	&3.33 \\
			&GMM-SENet(2P2S)	&0.0115	&0.35	&0.0826	&3.10 \\
			
			\hline
		\end{tabular}
	\end{center}
\end{table}

The proposed GMM-ResNet and GMM-SENet models outperform the baseline system obviously. If we use the ResNet and SENet to model the LFCC and CQCC features without GMM, the system performance decreases. So, the role of GMM is important, and this also proves the advantage of the GMM feature. The two-path architecture and two-step training scheme can further improve performances. The models with LFCC feature are obviously better than that with CQCC feature. Compared with the GMM, the LFCC+GMM-ResNet(2P2S) system can relatively reduce min-tDCF and EER by 76.1\% and 76.3\% on the evaluation dataset.

In order to compare with other systems, a simple linear function is used to combine the scores of the sub-systems in the experiments. The weights of sub-systems are only trained on the development set. Table \ref{tab:asvspoof2019la_fusion} shows the results. After score fusion, the min-tDCF and EER are 0, and this maybe mean that the fusion system is overfitting on the development set. So, the weight parameters in fusion function are not uniquely determined. If we change the weights slightly, the EER and min-tDCF are still 0 on development set, but them varies significantly on evaluation set. Table \ref{tab:asvspoof2019la_fusion} shows the typical fusion results, and we can’t obtain the best performance of fusion system on the evaluation set. Moreover, the performance of LFCC+GMM-ResNet is second only to T05 system on the leaderboard of the ASVspoof 2019 challenge, even though these competing systems use an ensemble of classifiers and neural networks in either the front- or back-end. On the other hand, the proposed LFCC+GMM-ResNet system achieves a superior performance compared with the state-of-the-art single systems, such as the LFCC+LCNN-LSTM\cite{wang21fa_interspeech} and CQT+SE-Res2Net50+CE\cite{li_replay_2021} systems.

\begin{table}[t]\setlength\tabcolsep{3pt} 
	\begin{center}
		
		\caption{Performance comparison of our proposed system to the top-performing systems (T05 and T45) and the state-of-the-art single systems on the ASVspoof 2019 LA scenario.} \label{tab:asvspoof2019la_fusion}

		\begin{tabular}{clcccc}
			\hline
			\multirow{2}*{Feature} & \multirow{2}*{Model}  & \multicolumn{2}{c}{Dev}  & \multicolumn{2}{c}{Eval} \\
			\cline{3-6}
			&	& tDCF & EER  & tDCF & EER \\
			\hline
			\multirow{2}*{CQCC} 
			&GMM-ResNet	        &0.0024	 &0.09	&0.2039	&7.83 \\
			&GMM-SENet       	&0.0001  &0.01	&0.2000	&7.67 \\
			\hline
			\multirow{2}*{LFCC} 
			&GMM-ResNet	        &0.0094	 &0.35	&\textbf{0.0498}	&\textbf{1.80} \\
			&GMM-SENet       	&0.0115  &0.35	&0.0826	&3.10 \\
			\hline
			\multicolumn{2}{c}{Fusion of the above four systems} &\textbf{0.0000}	&\textbf{0.00}	&0.0466	&1.86  \\
			\multicolumn{2}{c}{T05 \cite{todisco19_interspeech}} &-	 &-	&0.0069	&0.22  \\
			\multicolumn{2}{c}{T45 \cite{todisco19_interspeech}} &-	 &-	&0.0510	&1.86  \\
			\multicolumn{2}{c}{LFCC+LCNN-LSTM \cite{wang21fa_interspeech}} &-	 &-	&0.052 	&1.92   \\
			\multicolumn{2}{c}{CQT+SE-Res2Net50+CE \cite{li_replay_2021}} &0.0143	 &0.432	&0.0743 	&2.502   \\
			\hline
		\end{tabular}
	\end{center}
\end{table}
\subsection{Results on ASVspoof 2019 PA scenario}
Spoofing attacks in physical access scenario are referred to replay attacks, whereby a recording of a bona fide access attempt is first captured, presumably surreptitiously, before being replayed to the ASV microphone. Table \ref{tab:asvspoof2019pa} shows the results on the ASVspoof 2019 PA scenario obtained by the baseline systems and our models with LFCC and CQCC.
\begin{table}[t]\setlength\tabcolsep{3pt} 
	\begin{center}
		\caption{Results on ASVspoof 2019 PA scenario in terms of minimum tDCF and EER (\%). } \label{tab:asvspoof2019pa}
		\begin{tabular}{clcccc}
			\hline
			\multirow{2}*{Feature} & \multirow{2}*{Model}  & \multicolumn{2}{c}{Dev}  & \multicolumn{2}{c}{Eval} \\
			\cline{3-6}
			&	& tDCF & EER  & tDCF & EER \\
			\hline
			\multirow{8}*{CQCC}&GMM \cite{todisco19_interspeech} & 0.1953 & 9.87 & 0.2454 & 11.04 \\
			&GMM(ours)          &0.1843  &9.72  &0.2540 &11.34 \\ 
			\cline{2-6}
			&ResNet(without GMM)	        &0.0410	&1.63	&0.0505	&1.93 \\
			&GMM-ResNet	        &0.0281	&1.10	&0.0390	&1.48 \\
			&GMM-ResNet(2P)  	&0.0305	&1.24	&0.0392	&1.48 \\
			&GMM-ResNet(2P2S)	&0.0324	&1.26	&0.0380	&1.44 \\  
			\cline{2-6}
			&SENet(without GMM)       	&0.0485	&1.87	&0.0663	&2.49 \\
			&GMM-SENet       	&0.0266	&1.02	&0.0401	&1.47 \\
			&GMM-SENet(2P)   	&0.0301	&1.09	&0.0392	&1.49 \\
			&GMM-SENet(2P2S)	&0.0244	&0.98	&\textbf{0.0336}	&\textbf{1.27} \\
			\hline
			\multirow{8}*{LFCC}&GMM\cite{todisco19_interspeech} & 0.2554	&11.96	&0.3017	&13.54 \\
			&GMM(ours)          &0.2363	&11.22	&0.2869	&12.90 \\  
			\cline{2-6}
			&ResNet(without GMM)	        &0.0254	&0.89	&0.0357	&1.29 \\
			&GMM-ResNet	        &0.0232	&0.81	&0.0293	&0.99 \\
			&GMM-ResNet(2P)   	&0.0226	&0.80	&0.0257	&0.93 \\
			&GMM-ResNet(2P2S)	&0.0237	&0.81	&0.0262	&0.91 \\
			&SENet(without GMM)       	&0.0234	&0.87	&0.0307	&1.13 \\  \cline{2-6}
			&GMM-SENet	        &0.0182	&0.61	&0.0195	&0.69 \\
			&GMM-SENet(2P)   	&0.0148	&0.61	&0.0187	&0.65 \\
			&GMM-SENet(2P2S)	&0.0176	&0.67	&\textbf{0.0162}	&\textbf{0.59} \\
			
			\hline
		\end{tabular}
	\end{center}
\end{table}
Compared with the LA scenario, the performance of our models on PA scenario is significantly better. Similarly, the two-path architecture and two-step training scheme can further improve performances, and our models can obtain better performance using LFCC than CQCC. Specifically, the LFCC+GMM-SENet(2P2S) system reduce the min-tDCF and EER by 94.4\% and 95.4\% on the evaluation dataset compared with the baseline.

\begin{table}[t]\setlength\tabcolsep{3pt} 
	\begin{center}
		\caption{Performance comparison of our proposed system to the top-performing systems (T28 and T45) and the state-of-the-art single systems on the ASVspoof 2019 PA scenario.}
		 \label{tab:asvspoof2019pa_fusion}
		\begin{tabular}{clcccc}
						\hline
			\multirow{2}*{Feature} & \multirow{2}*{Model}  & \multicolumn{2}{c}{Dev}  & \multicolumn{2}{c}{Eval} \\
			\cline{3-6}
			&	& tDCF & EER  & tDCF & EER \\
			\hline
			\multirow{2}*{CQCC} 
			&GMM-ResNet	        &0.0324	 &1.26	&0.0380	&1.44 \\
			&GMM-SENet       	&0.0244  &0.98	&0.0336	&1.27 \\
			\hline
			\multirow{2}*{LFCC} 
			&GMM-ResNet	        &0.0237	 &0.81	&0.0262	&0.91 \\
			&GMM-SENet       	&0.0176  &0.67	&0.0162	&0.59 \\
			\hline
			\multicolumn{2}{c}{Fusion of the above four systems} &0.0097	 &0.41	&\textbf{0.0120}	&\textbf{0.43}  \\
			\multicolumn{2}{c}{T28 \cite{todisco19_interspeech}}  &-	 &-	&0.0096	&0.39  \\
			\multicolumn{2}{c}{T45 \cite{todisco19_interspeech}}  &-	 &-	&0.0122	&0.54  \\
			\multicolumn{2}{c}{CQT+SE-Res2Net50+CE \cite{li_replay_2021}}  &0.0086	 &0.329	&0.0116	&0.459  \\
			\hline
		\end{tabular}
	\end{center}
\end{table}

Table \ref{tab:asvspoof2019pa_fusion} shows the results of our score fusion system and the top-performing systems in the ASVspoof 2019 challenge. The fusion system gets the further performance improvement, and achieves performance second only to the best system (T28) on the leaderboard. Our system is also compared with the state-of-the-art single systems , and the system performance is slightly lower than the CQT+SE-Res2Net50+CE system \cite{li_replay_2021} performance.

\section{CONCLUSIONS}
In this paper, we proposed the two-path GMM-ResNet and GMM-SENet models for spoofing speech detection. The classical GMM accumulates the scores on all frames independently, and does not consider the contribution of each Gaussian component to the final score. And the relationship between adjacent frames is also been ignored. For an utterance, the Gaussian probability feature includes the score distribution on each GMM component. The ResNet and SENet are applied for spoofing detection with the two-path architecture and the two-step training scheme. The experimental results on the ASVspoof 2019 database show that the proposed two-path neural networks can improve performance greatly. In future, we will explore new network architecture to model LGP feature. On the other hand, they will also be applied to speaker recognition.

\section{ACKNOWLEDGEMENTS}
This work is supported by National Natural Science Foundation of P.R.China (62067004).

\bibliographystyle{IEEEbib}
\bibliography{refs}
\end{document}